\documentclass[letter]{aa} 

\usepackage{graphicx}
\usepackage{txfonts}
\usepackage[flushleft]{threeparttable} 
\usepackage{soul}

\usepackage{hyperref}
\hypersetup{colorlinks=true,linkcolor=blue,citecolor=blue,filecolor=blue,urlcolor=blue}

\newcommand{\ha}{H$\alpha$}

\newcommand{\hei}{He~{\sc i}}
\newcommand{\heii}{He~{\sc ii}}

\def\lesssim{\mathrel{\hbox{\rlap{\hbox{\lower4pt\hbox{$\sim$}}}\hbox{$<$}}}}

\def\gtrsim{\mathrel{\hbox{\rlap{\hbox{\lower4pt\hbox{$\sim$}}}\hbox{$>$}}}}

\begin{document} 

   \title{Near-infrared emission lines trace the state-independent accretion disc wind of the black hole transient MAXI~J1820+070}

   \author{J. Sánchez-Sierras
          \inst{\ref{i1},\ref{i2}}
          \and
          T. Muñoz-Darias\inst{\ref{i1},\ref{i2}}
          }

   \institute{Instituto de Astrofísica de Canarias, E-38205 La Laguna, Tenerife, Spain \label{i1}
         \and
             Departamento de Astrofísica, Universidad de La Laguna, E-38206 La Laguna, Tenerife, Spain \label{i2}
             }

   \date{Accepted 9 Jul, 2020}
\titlerunning{near-infrared winds in MAXI~J1820+070}
\authorrunning{Sánchez-Sierras \& Muñoz-Darias}

  \abstract
  {The black hole transient MAXI J1820+070 displayed optical P-Cyg profiles and other wind-related emission line features during the hard state of its discovery outburst. 
  We present near-infrared (NIR) spectroscopy covering the different accretion states of the system during this event. Our eight-epoch data set (VLT/X-shooter) reveals strong variability in the properties of the NIR emission lines. This includes absorption troughs and extended emission line wings with kinetic properties that are remarkably similar to those inferred from the wind signatures observed in optical emission lines, indicating that they most likely trace the same accretion disc wind.
  Unlike the optical features, these NIR signatures are not exclusive of the hard state, as they are also witnessed across the soft state with similar observational properties.
  This supports the presence of a relatively steady outflow during the entire outburst of the system, and it represents the first detection of an accretion disc wind in a black hole soft state at energies other than X-rays. We discuss the visibility of the wind as a function of the spectral band and the potential of NIR spectroscopy for wind studies, in particular during luminous accretion phases.}


   \maketitle

\section{Introduction}
\defcitealias{Munoz-Darias2019}{MD19} 

\begin{table*}[ht!]
        \centering
        \caption{Observing epochs, accretion state, and wind behaviour.}
        
        \begin{threeparttable}
        \begin{tabular}{l c c c c}
        
                \hline
                Epoch [\citetalias{Munoz-Darias2019}] & MJD (date) & Accretion state & Optical winds\tnote{a,b} & NIR winds\tnote{b}
                \\
                \hline
                \hline
                1 [2] & 58193 (2018-03-16) & Hard & P-Cyg & AT(\hei--10830) \\
                2 [6] & 58197 (2018-03-20) & Hard & BW & BW \\
                3 [9] & 58199 (2018-03-22) & Hard & BW & BW/AT \\
                4 [22] & 58312 (2018-07-13) & Soft & -- & BW/AT \\
                5 [32] & 58368 (2018-09-07) & Soft & -- & BW/AT \\
                6 [33] & 58390 (2018-09-28) & Hard-intermediate & P-Cyg/BW & BW \\
                7 [34] & 58391 (2018-09-29) & Hard-intermediate & P-Cyg & BW \\
                8 [--] & 58559 (2019-03-17) & Hard & P-Cyg & AT(\hei--10830) \\
                \hline
        \end{tabular}
        \label{dataTable}
    \begin{tablenotes}
        \item[a] From \citetalias{Munoz-Darias2019}.  
        \item[b] Wind signatures: P-Cyg profiles, broad emission line wings (BWs), and absorption troughs (ATs).
    \end{tablenotes}
        \end{threeparttable}
\end{table*}

The outbursts of transient low-mass X-ray binaries (LMXBs) produce large amounts of radiation from X-rays to radio waves. These events allow us to study the complex balance between accretion and its associated outflows in great detail and across a wide range of luminosities (e.g. \citealt{Fender2016}). Accretion disc winds are arguably the last relevant ingredient added to the LMXB accretion picture, since there is increasing evidence suggesting that they carry away a significant fraction of the mass involved in the accretion process (e.g. \citealt{Neilsen2011,Ponti2012,Munoz-Darias2016,Casares2019}) as well as angular momentum (e.g. \citealt{Tetarenko2018, Dubus2019}). Winds are detected in two flavours, the so-called hot and cold winds. The former are revealed by X-ray spectroscopy through the blue-shifted absorption lines of mildly to highly ionised species (e.g. \citealt{Miller2006}) that have been detected in high inclination sources (\citealt{Ponti2016, DiazTrigo2016}). Cold winds are detected in optical emission lines of (mainly) helium (e.g. \hei\ at 5876 \AA; \hei--5876 hereafter) and hydrogen (Balmer series, especially \ha) as P-Cyg profiles as well as broad emission line wings, which are typically superimposed on the accretion disc emission lines. They have been discovered in several black hole (BH) transients [e.g. \citealt{Munoz-Darias2016,Munoz-Darias2018, Charles2019,Jimenez-Ibarra2019b};  \citealt{Munoz-Darias2019}, hereafter \citetalias{Munoz-Darias2019}] as well as in Swift J1858.6-0814, which most likely harbours a neutron star \citep{Munoz-Darias2020, Buisson2020}.

The coupling between the properties of the outflow and those of the accretion flow is robustly established for jets (see e.g. \citealt{Fender2004}). However, from a purely observational point of view, the situation is more complex for winds. Hot X-ray winds are preferentially seen during soft X-ray states (i.e. when the jet is quenched; \citealt{Neilsen2009,Ponti2012}),  whereas optical winds have been observed so far in harder states characterised by continuous (radio) jet emission, albeit some of the systems displaying optical winds do not follow a standard outburst evolution (see e.g. \citealt{McClintock2006, Done2007, Belloni2011} for reviews on BH accretion states). The canonical behaviour, with long-lived hard and soft states \citep{Shidatsu2019}, of the BH transient MAXI~J1820+070 (ASASSN-18ey, \citealt{Tucker2018}; hereafter J1820) during its 2018 outburst offered a great opportunity to study this topic. An extensive spectroscopic campaign revealed optical wind signatures during the hard state that were not observed in the soft state (\citetalias{Munoz-Darias2019}). This behaviour was suggested to be linked to a change in the ionisation state of the ejecta, as it was observed in the BH transient V404 Cyg, where the visibility of the optical wind was found to be strongly affected by ionisation effects \citep[e.g.][]{MataSanchez2018}. In this letter, we present eight epochs of high-quality near-infrared (NIR) spectroscopy supporting the presence of a cold accretion disc wind during both the hard and soft states of this dynamically confirmed \citep{Torres2019,Torres2020} BH transient.


\begin{figure}
    \centering
    \includegraphics[width=9truecm]{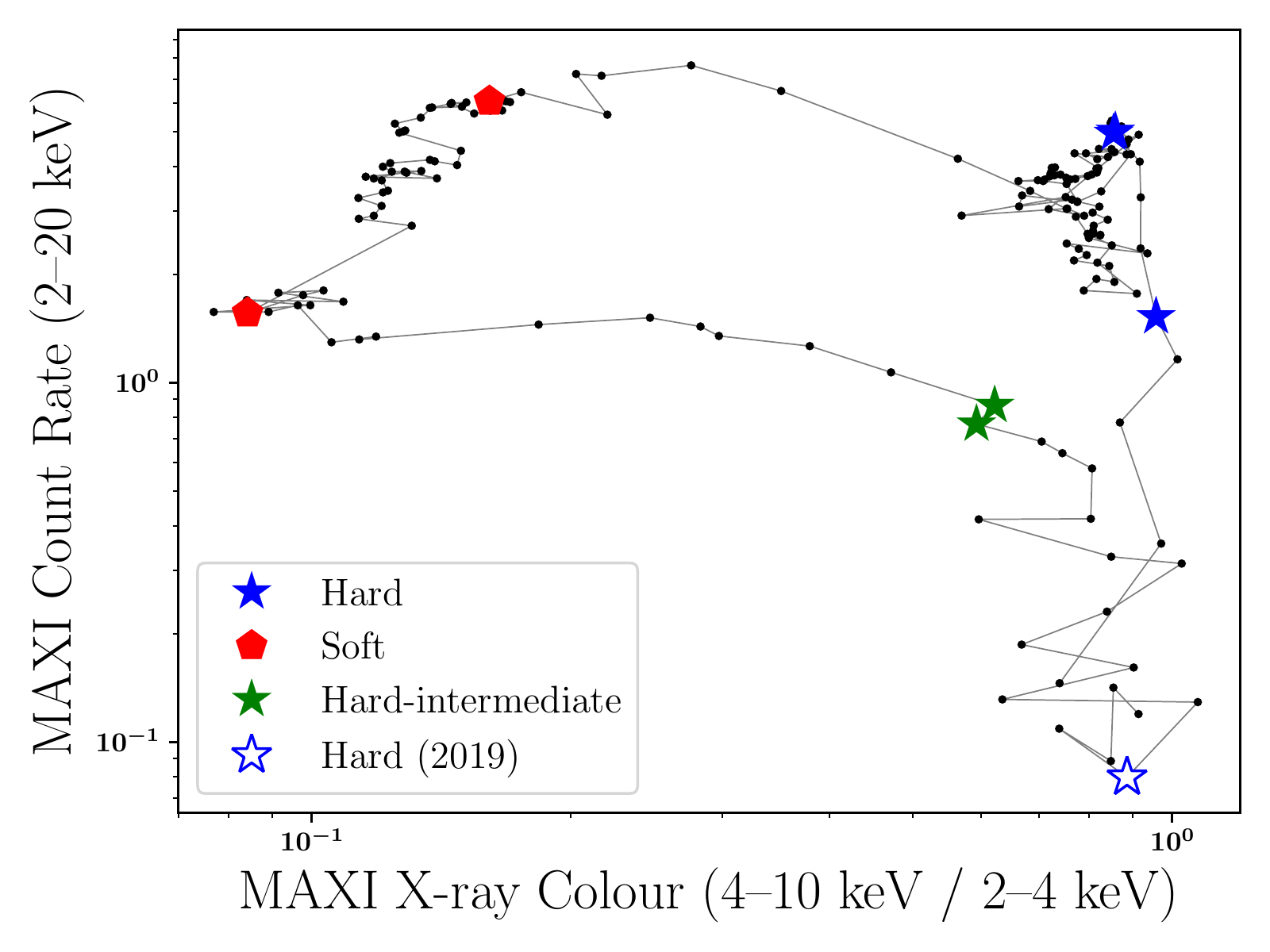}
    \caption{Hardness-intensity diagram of J1820 (from MAXI). The eight NIR spectroscopic epochs are marked as indicated in the legend. Epochs 2 and 3 overlap. Epoch 8 (empty symbol) was taken during a re-brightening in 2019 (i.e. 169 days apart).}
    \label{figHID}
\end{figure}

\begin{figure*}
    \centering
    \includegraphics[width=18truecm]{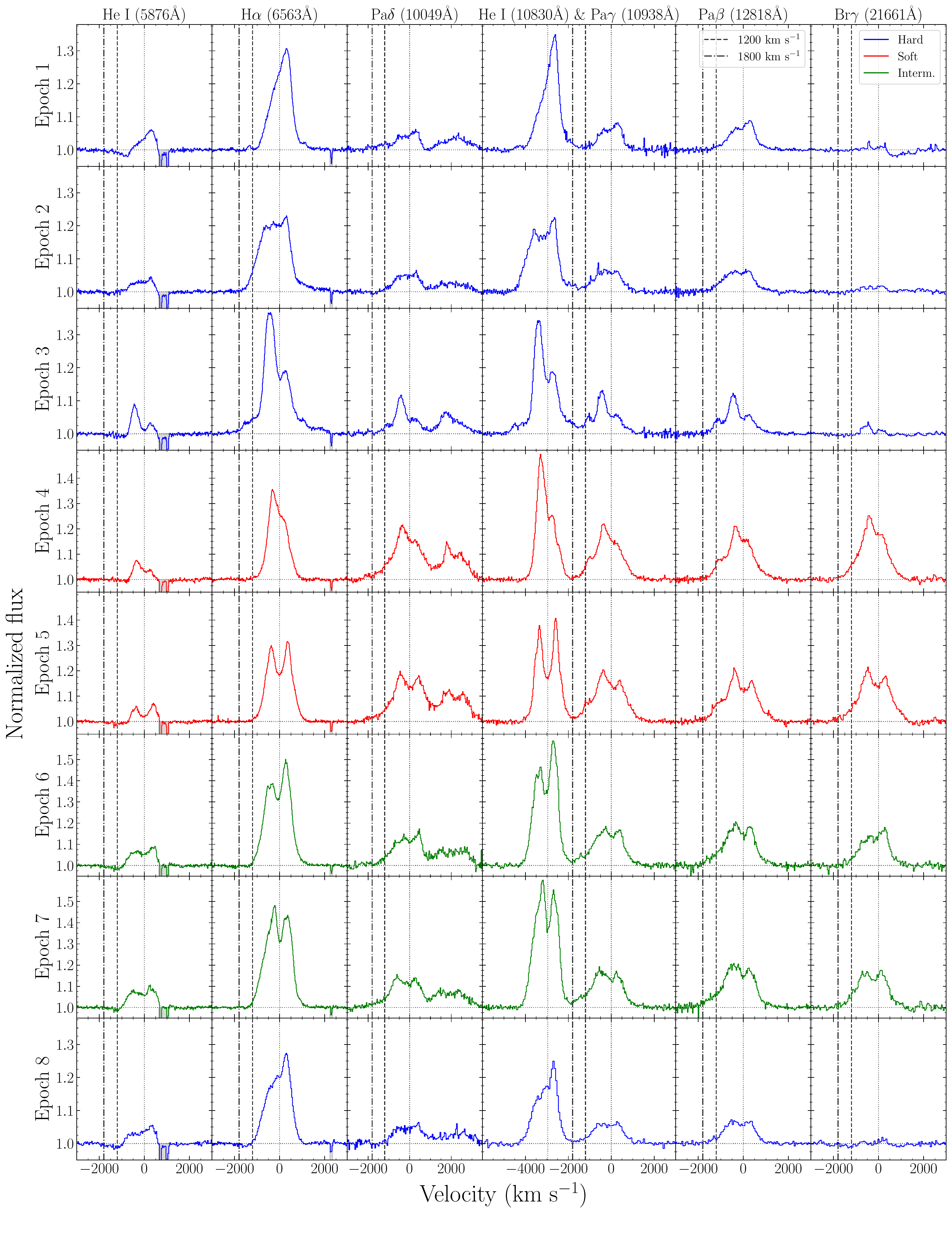}
    \caption{Evolution of the main NIR emission lines across the outburst. The optical transitions \ha~and \hei--5876 lines are also included for comparison. Every line is plotted normalised in a separate panel, with the exception of \hei--10830 and Pa$\gamma$, which share axes due to their proximity. \heii~at 10124~\AA\ can also be seen in the Pa$\delta$ panels, especially during the soft-state epochs. The characteristic wind velocities determined in \citetalias{Munoz-Darias2019} are plotted as dashed (1200 km s$^{-1}$) and dash-dotted (1800 km s$^{-1}$) lines.}
    \label{figShouldersCompare}
\end{figure*}
\section{Observations and data reduction}
Out of the 37 epochs of optical spectroscopy presented in \citetalias{Munoz-Darias2019}, 7 epochs correspond to observations performed with the X-shooter spectrograph \citep{Vernet2011} attached to the Very Large Telescope (VLT). This instrument has a NIR arm (10000--24800~\AA), which is operated simultaneously with the visible and ultraviolet arms (whose data were included in \citetalias{Munoz-Darias2019}). We focus our study on a detailed analysis of this NIR spectroscopy. Furthermore, we present an additional X-shooter epoch taken during a re-brightening in March 2019 \citep{Bahramian2019}. The optical data from this observation were obtained and analysed in the same way as those presented in \citetalias{Munoz-Darias2019}. This NIR spectrum was added to the data set for a total of 8 NIR spectroscopic epochs (see Table \ref{dataTable}).  

We obtained between 8 and 12 exposures per epoch using an ABBA nodding configuration, with total exposure times in the range of 1160--2160 s. We used a slit width of 0.9 arcsec, which rendered a velocity resolution of $\sim$ 54 km s$^{-1}$. The data were reduced using the X-shooter ESO Pipeline v3.3.5. Thus, they were extracted into mono-dimensional spectra that are wavelength and flux calibrated. All the spectra were carefully corrected for telluric absorption lines using \textsc{Molecfit} \citep{Smette2015}. The data analysis was performed using \textsc{Molly} and custom routines developed under \textsc{Python} 3.7.

\section{Results and analysis}
\label{results}

The discovery outburst of J1820 was first detected in X-rays in March 2018 \citep{Kawamuro2018} by the MAXI instrument \citep{Matsuoka2009} on board the International Space Station. It showed both hard and soft states (see Fig. \ref{figHID}) that trace an anticlockwise hysteresis pattern in the hardness-intensity diagram  (\citealt{Homan2001}). Epochs 1--3 were taken during the initial hard state within 11 days from the outburst detection (see Table \ref{dataTable}), while epochs 4 and 5 (4--6 months later) correspond to the slow decay of the system during the soft state. In addition, two more epochs were taken towards the end of the soft-to-hard transition. According to the hardness-intensity and rms-intensity (\citealt{Munoz-Darias2011}) diagrams presented in \citet{Stiele2020}, they correspond to the hard-intermediate state. Finally, epoch 8 (hard state) was obtained in 2019 during a re-brightening phase at lower luminosity than those from the 2018 event.

Fig. \ref{figShouldersCompare} shows for each epoch the strongest NIR emission lines in the spectrum. These include \hei~at 10830\AA~(\hei--10830 hereafter), as well as hydrogen lines corresponding to the Paschen (Pa$\delta$, Pa$\gamma$, Pa$\beta$) and Brackett (Br$\gamma$) series. The optical lines showing the strongest wind features  (H$\alpha$ and \hei--5876; \citetalias{Munoz-Darias2019}) are also shown for comparison. We note that other NIR emission lines are sometimes detected, such as higher order transitions of the Bracket series. However, besides being weaker, these are clustered on a relatively small spectral region, not allowing for a proper continuum determination. Furthermore, \heii~at 10124~\AA\ (i.e. redward of Pa$\delta$) is also detected, especially during the soft state epochs. This line has not been included in our analysis given its weakness.  

Fig. \ref{figShouldersCompare} reveals the striking similarity between H$\alpha$ and \hei--10830 in every epoch. In addition, the reddest transition, Br$\gamma$, is not present or just marginally detected during the four hard-state epochs (see section \ref{discussion}). The remaining emission lines are always observed and display remarkable variability during the outburst. This includes standard double-peaked profiles (e.g. \citealt{Smak1969}), but also additional features that can be interpreted as accretion disc wind signatures (see below). In the blue wing of each line in Fig. \ref{figShouldersCompare} we mark with dashed and dash-dotted vertical lines the two wind terminal velocities (1200 and 1800 km s$^{-1}$, respectively) determined by \citetalias{Munoz-Darias2019}. They correspond to the blue edge of the P-Cyg absorption components that were detected throughout the outburst. The highest velocity also matches that of the edges of highly significant broad emission line wings that were detected at different epochs, in some cases, simultaneously with the P-Cyg profiles (see fig 2 in \citetalias{Munoz-Darias2019}).

\subsection{Hard state: conspicuous optical and NIR wind signatures}

Epoch 1 shows a P-Cyg profile in \hei--5876 with a blue-edge velocity of $\sim 1200$ km s$^{-1}$ (see also \citetalias{Munoz-Darias2019}). In the NIR, the Paschen transitions are asymmetric (as well as \ha\ and \hei--10830), with Pa$\delta$ displaying a blue emission line wing that reaches $ \sim 1800$ km s$^{-1}$. This extended wing also shows a hint of an absorption trough (see below) with a similar blue-edge velocity as that of the \hei--5876 P-Cyg profile (i.e. $\sim 1200$ km s$^{-1}$; see the top left panel in Fig \ref{figShouldersDetail}).

Epochs 2 and 3 were both taken at the hard-state peak with a separation of just two days. The former is characterised by relatively flat-top profiles in both the optical and NIR transitions, and \ha~shows an extended red-wing component (see also \citetalias{Munoz-Darias2019}). Two days later (epoch 3), the system displayed a remarkable optical-NIR spectrum. \ha\ (and \hei--10830) shows an underlying broad component that extends up to $\pm 1800$ km s$^{-1}$. This was also detected by \textit{Gran Telescopio Canarias} (GTC) observations performed 25 hours before (see fig. 2 in \citetalias{Munoz-Darias2019}). The broad underlying component is also present in the NIR transitions. In addition, the Paschen lines show an absorption trough with a blue-edge velocity of $\sim 1200$ km s$^{-1}$ (dashed line in Fig. \ref{figShouldersCompare}; see also the top left panel in Fig. \ref{figShouldersDetail}). This was also present in the \ha\ line profile observed by GTC the night before, when a P-Cyg profile indicating the same wind terminal velocity was present in \hei--5876 and \hei--6678 (see fig. 1 and 2 in \citetalias{Munoz-Darias2019}). Thus, we interpret the absorption troughs witnessed in the Paschen lines as NIR signatures of the accretion disc wind that is detected at optical wavelengths.

\subsection{Soft-state NIR winds}
Epoch 4 corresponds to the most luminous phase of the soft state, while epoch 5 (56 days later; see Table \ref{dataTable}) was taken at lower luminosity and close to the soft-to-hard transition. As reported in \citetalias{Munoz-Darias2019}, the optical wind features detected during the hard state are not present in these observations. However, some of the NIR signatures remain. In particular, Pa$\gamma$ and Pa$\beta$ show remarkable absorption troughs in the blue wing, with a blue-edge velocity of $-1200$ km s$^{-1}$ (i.e. consistent with that determined in several optical P-Cyg profiles during the hard state; \citetalias{Munoz-Darias2019}). Fig. \ref{figShouldersDetail} (top panels) shows how these troughs, that take the shape of emission line shoulders, are an evolution of those seen in the hard state in the very same NIR transitions. Likewise, when comparing Pa$\gamma$ and Pa$\beta$ with Br$\gamma$, which shows a rather standard double-peaked profile, the blue wings of the Paschen lines appear to be absorbed for velocities lower than 1200 km s$^{-1}$. This is precisely the blue-edge velocity of the P-Cyg profiles and absorption troughs observed in the hard state (Fig. \ref{figShouldersDetail}, bottom left panel). All these observables strongly suggest that during the soft state, the Paschen NIR emission lines are still sensitive to the presence of the accretion disc wind observed during the hard state. 

Epochs 6 and 7 were taken on two consecutive nights 22 days later, towards the end of the soft-to-hard transition during the hard-intermediate state. \citetalias{Munoz-Darias2019} (bottom panel in fig. 1) reported the P-Cyg profile detected in \hei--5876, which was particularly conspicuous in Epoch 6, with a blue-edge velocity of $\sim$ 1800 km s$^{-1}$. In the NIR, we no longer detect the blue absorption troughs observed in epochs 3, 4, and 5 (but see below).

\subsection{Wind detection during the 2019 low-luminosity hard state}
\begin{figure*}[!ht]
    \centering
    \includegraphics[width=18truecm]{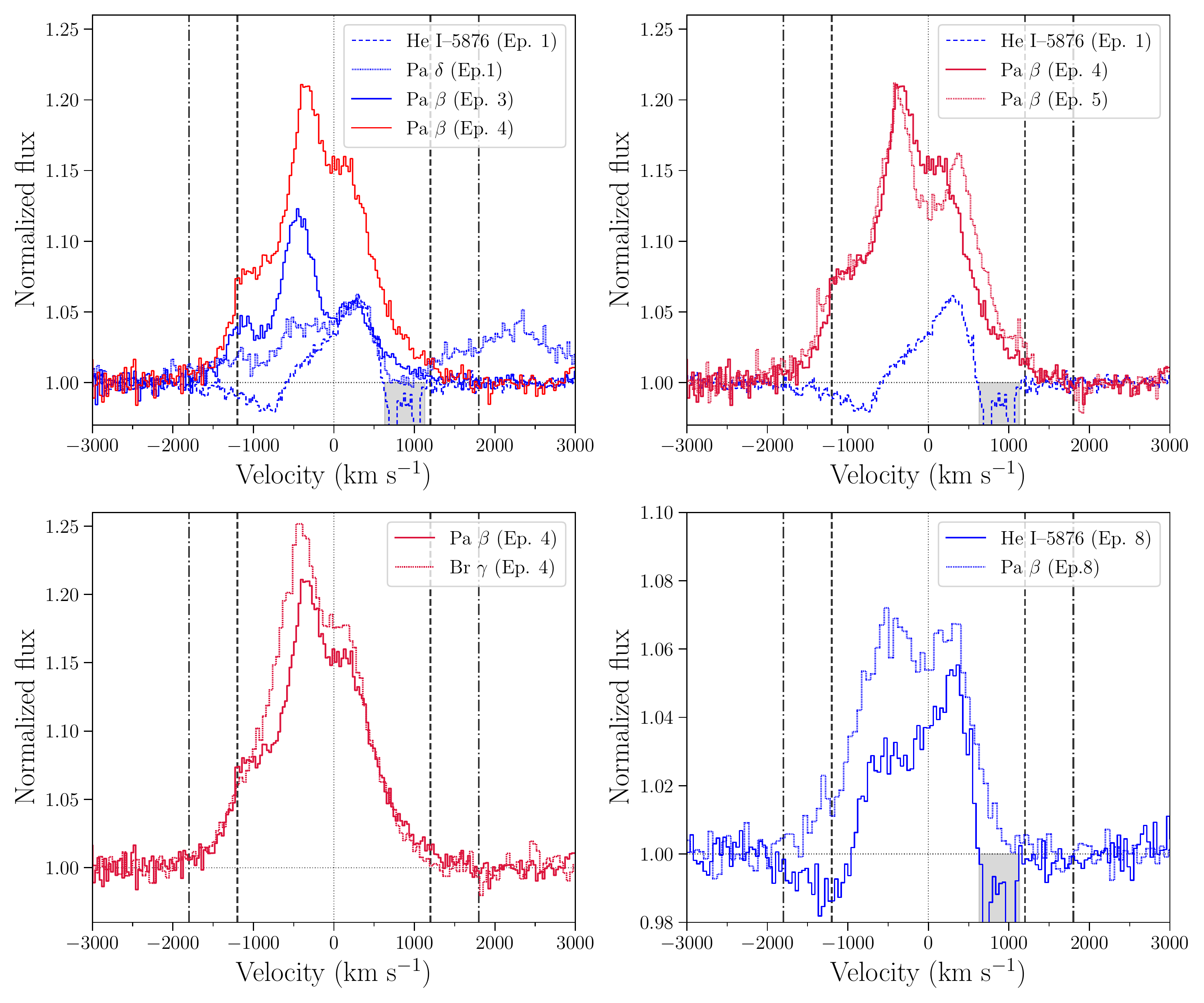}
    \caption{Detailed comparison of some of the emission lines presented in Fig. \ref{figShouldersCompare}. The characteristic wind velocities (from \citetalias{Munoz-Darias2019}) are marked as dashed ($\pm 1200$ km s$^{-1}$) and dash-dotted ($\pm 1800$ km s$^{-1}$) vertical lines. The colour code reflects the different X-ray states as in Fig. \ref{figShouldersCompare}. The Na interstellar doublet in the red wing of \hei--5876 is shaded. \textit{Top left panel}: Pa$\delta$ (epoch 1) and Pa$\beta$ (epochs 3 and 4) during the rising phase of the outburst. Blue-shifted absorption troughs with a blue-edge velocity of $\sim 1200$ km s$^{-1}$ are visible. These are compared with the \hei--5876 P-Cyg profile from epoch 1. The \textit{top right panel} shows the two soft-state epochs (4 and 5) of Pa$\beta$ against the \hei--5876 P-Cyg profile from epoch 1. \textit{Bottom left panel:} Pa$\beta$ and Br$\gamma$ from epoch 4 showing clear differences in the blue part of the line up to $\sim 1200$ km s$^{-1}$. \textit{Bottom right panel:} \hei--5876 (epoch 8) showing a P-Cyg profile with a terminal velocity of $\sim 1800$ km s$^{-1}$. Pa$\beta$ (also epoch 8) displays a double-peaked profile whose blue wing reaches higher velocities ($\sim 1800$ km s$^{-1}$) than the red wing. The flux scale in this panel is different from that of the previous panels.}
    \label{figShouldersDetail}
\end{figure*}

An additional epoch (low-luminosity hard state), that was not included in \citetalias{Munoz-Darias2019}, was taken during a brief re-brightening phase while the system was returning to quiescence  (epoch 8; see Table \ref{dataTable} and Fig. \ref{figHID}). \hei--5876 is skewed towards the red and again displays a weak but clearly detected P-Cyg profile with a blue-edge velocity of $\sim 1800$ km s$^{-1}$. The core of the absorption component reaches $\sim 2\%$ below the continuum level, consistent with the P-Cyg profiles that were detected during the initial hard state. \ha\ and \hei--10830 are also skewed towards the red, probably as a result of the same outflowing material that was detected in \hei--5876. The Paschen lines do not display absorption troughs, but instead show (as in epochs 6 and 7) blue wings that extend (rather precisely) up to $\sim 1800$ km s$^{-1}$ (bottom right panel in Fig. \ref{figShouldersDetail}). This suggests that they might still be sensitive to the outflow. However, a symmetrically broad red wing such as that observed in \ha\ (and \hei--10830) in epoch 3 might be expected in this case.

\begin{figure}[!ht]
    \centering
    \includegraphics[width=9truecm]{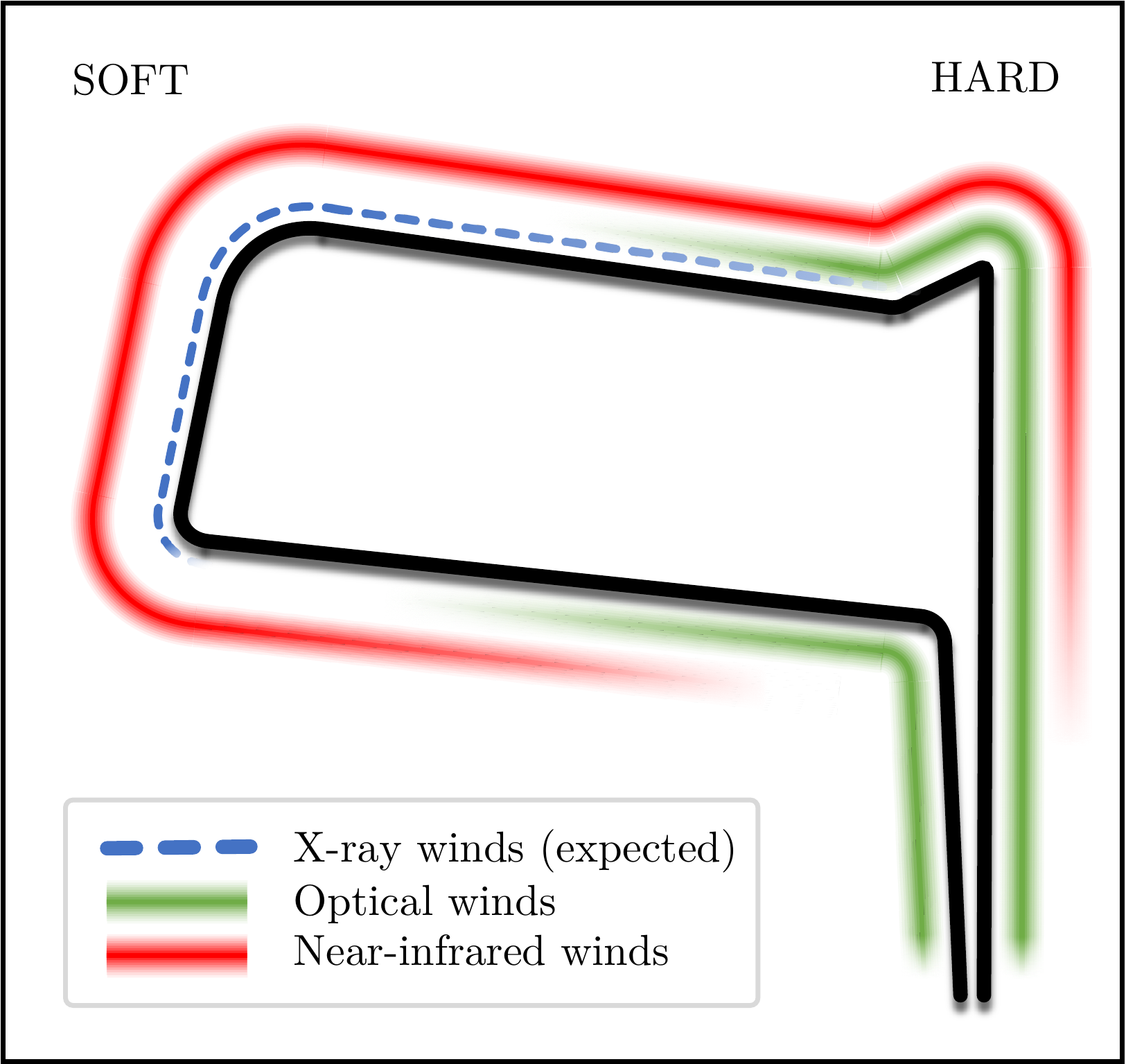}
    \caption{Sketch of the wind visibility across the hardness-intensity diagram of MAXI J1820+070 (based on \citetalias{Munoz-Darias2019} and this work). The blurred lines indicate the (approximate) location of the wind detections at optical and NIR wavelengths. The outburst stages that are more favourable for detecting X-ray winds (based on \citealt{Ponti2012}) are marked by a dashed blue line.}
    \label{figSketch}
\end{figure}

\section{Discussion}
\label{discussion}

We have presented NIR spectroscopic observations sampling the entire 2018--2019 outburst of the BH transient J1820. Our eight-epoch data set  reveals the presence of blue-shifted absorption troughs and conspicuous broad emission line wings in some of the most prominent NIR emission lines. The kinetic properties of these features are very similar to those indicated by the P-Cyg profiles detected in optical lines (see the top left panel in Fig \ref{figShouldersDetail}). In addition, similar absorption troughs and broad emission components superimposed on standard double-peaked profiles were also detected in optical lines, sometimes simultaneously with P-Cyg profiles (see also \citetalias{Munoz-Darias2019}). All the above strongly suggests that both the NIR and the optical emission lines trace the same cold accretion disc wind. However, unlike the optical wind signatures, the NIR features are not exclusive to the hard state and are also detected in the soft state. A sketch that broadly summarises the visibility of the wind across the outburst of J1820 and as a function of the spectral band is presented in Fig. \ref{figSketch}. We also mark the (approximate) expected location of X-ray winds based on \citet{Ponti2012}. We note that there is some evidence indicating that the orbital inclination of J1820 is relatively high (e.g. \citealt{Kajava2019, Torres2020, Bright2020}), which would favour the detection of hot winds. However, the system was not observed by \textit{XMM-Newton} or \textit{Chandra} during the soft state.  

The number of LMXBs that have shown NIR features that are consistent with  disc winds is very limited. The nebular phase \citep[strong and broad emission line components produced by a massive wind; see][]{Munoz-Darias2016,Munoz-Darias2017,Casares2019} following the luminous 1999 outburst peak of the BH transient V4641 Sgr \citep{Munoz-Darias2018} was also detected in Br$\gamma$ \citep{Chaty2003}. In addition, \citet{Rahoui2014} observed a slightly blue-shifted broad emission component underlying Pa$\beta$ during the low-luminosity hard state of GX~339-4. 
In neutron star LMXBs, \citet{Bandyopadhyay1999} detected a Br$\gamma$ P-Cyg profile in GX 13+1 (and possibly in Sco X-1; but see \citealt{MataSanchez2015, Homan2016}). The exact accretion phase corresponding to these detections is unknown, but the X-ray properties of these persistently luminous sources (known as Z-sources; e.g. \citealt{vanderKlis2006}) resemble to some extent those of BHs in intermediate and soft states at high luminosity \citep{Munoz-Darias2014}. We note that this low number of wind detections in the NIR likely arises because only a few sensitive spectroscopic campaigns have been carried out to date and do not imply that wind signatures in the NIR are rare.

\subsection{Steady, state-independent NIR wind traced by the Paschen lines}

The NIR spectra show that the optical-NIR wind observed during the hard state is also present in the soft state. Although the state classification and outburst evolution is not trivial for the sources displaying the strongest cold wind signatures (see \citealt{Munoz-Darias2020}), these results represent the first detection of an accretion disc wind in a steady BH soft state at energies other than X-rays. 

The Pa$\gamma$ and Pa$\beta$ blue-shifted absorption troughs are clearly detected in the bright hard-state (epoch 3) and the two soft-state epochs taken more than 100 days later (epochs 4 and 5). These are separated by 56 days and a factor of four in X-ray luminosity (see Fig. \ref{figHID}). However, the absorption troughs have a comparable strength and indicate similar kinetic properties during (at least) the most luminous phase of the outburst (see Fig. \ref{figShouldersDetail}). This strongly suggests the presence of a rather steady outflow throughout the different accretion states. The detection of a state-independent wind is in agreement with \citet{Tetarenko2018}, who invoked the presence of such an outflow to explain the unrealistically high levels of angular momentum removal (for a viscous disc) derived from fits to  X-ray light curves of BH transients.

The soft state of J1820 was characterised by stronger high-ionisation emission lines (\heii--4686 and the Bowen blend; see \citetalias{Munoz-Darias2019}). The effect of ionisation in the wind (P-Cyg profiles) of massive stars was modelled by \citet{Najarro1997}. They showed that an increase in the ionisation level might cause the blue-shifted Balmer absorption profiles to disappear, while hardly affecting those in the Paschen series. As the ejecta become more ionised, the higher energy levels (n=3 for Paschen) become more populated, which might explain why the NIR transitions are more sensitive to winds than the Balmer lines during the soft state. This might be also a viable explanation for the above-mentioned Br$\gamma$ P-Cyg profiles found in luminous neutron stars. Nevertheless, a significant caveat that arises when comparing massive stars with LMXBs is the high-energy emission produced by the latter group, which is expected to fully ionise the ejecta unless some shielding is present (e.g. \citealt{Proga2002}). In this regard, it is worth mentioning that at least for one system, the optical wind has been proposed to be clumpy \citep{Charles2019,Jimenez-Ibarra2019b}. In addition, we note that in addition to the changes in luminosity, the density of the ejecta and the variations in the spectral energy distribution of the radiation field must play a role in the wind detectability for the different spectral ranges (see e.g. \citealt{Chaty2003,Bianchi2017}).   
\subsection{\ha, \hei--10830 and Br$\gamma$ }

Fig. \ref{figShouldersCompare} shows that \ha~and \hei--10830 display virtually identical line profiles and a very similar behaviour across all the epochs. It is important to bear in mind that in both transitions the lower energy level [n=2 for H and the 2s (triplet) for He] is meta-stable. Our study shows that \hei--10830, which is only weakly affected by telluric absorptions, might be an interesting alternative to \ha~in order to detect cold winds in (for instance) extinguished objects.

At the opposite side of the NIR spectrum, Br$\gamma$ has not proved to be particularly sensitive to winds in J1820. In the first place, the line is very weak or not detected at all during the hard state, most likely because of the strong contribution to the NIR of the jet component. This is supported by \cite{Shidatsu2019}, who inferred a strong jet contribution to the optical (and therefore NIR) emission during this phase of the outburst (see also \citealt{Russell2006}). However, a strong Br$\gamma$ emission line is present during the soft and hard-intermediate states, when the jet component is expected to recede (or completely quench; e.g. \citealt{Russell2011}). It shows relatively broad wings in epochs 4 to 7 (Fig. \ref{figShouldersCompare}) and is skewed towards the red in epoch 6 (bottom left panel in Fig. \ref{figShouldersDetail}), but there is no definitive wind signature detected at any epoch. This is in contrast with the above-mentioned Br$\gamma$ P-Cyg profiles witnessed in luminous neutron stars. However, we note that the peak luminosity of the J1820 outburst was only $\sim$ 15\% of the Eddington luminosity \citep{Atri2020}, that is, significantly lower than typical for Z-sources.

\section{Conclusions}
We presented a NIR spectroscopic follow-up of the BH transient MAXI~J1820+070, in which an optical accretion disc wind had been observed during the hard state. We reported the presence of several NIR emission lines features, particularly blue-shifted absorption troughs, whose kinetic properties and time evolution strongly suggest that they are tracing the aforementioned cold outflow. The NIR wind signatures are found during both the hard and soft states, indicating that the accretion disc wind was most likely active during the entire outburst. This study represents the first detection of an accretion disc wind during the soft state at energies other than X-rays, showing the potential of NIR spectroscopy to study such outflows at high luminosities and during disc-dominated outburst phases.
\begin{acknowledgements}
We are thankful to the anonymous referee for constructive comments that have improved this letter. We acknowledge support by the Spanish Ministerio de Ciencia e Innovaci\'on under grant AYA2017-83216-P. TMD acknowledges support via the Ram\'on y Cajal Fellowships RYC-2015-18148. Based on observations collected at the European Southern Observatory under ESO programmes 0100.D-0292(A), 0101.D-0158(A) and 0102.D-0309(A). This research has made use of MAXI data provided by RIKEN, JAXA and the MAXI team. \textsc{Molly} software developed by Tom Marsh is gratefully acknowledged.
\end{acknowledgements}

\bibpunct{(}{)}{;}{a}{}{,} 
\bibliographystyle{aa}

\end{document}